\documentclass[preprint,12pt]{elsarticle}
\usepackage{amsmath}
\usepackage{epsf}
\usepackage{graphicx}

\begin{document}

\begin{frontmatter}

\title{Neutrino speed anomaly as signal of Lorentz violation$^\dagger$\footnote{$\dagger$Published as L.~Zhou and B.-Q.~Ma, Astropart.Phys. 44 (2013) 24-27.}}

\author[pku]{Zhou Lingli}\ead{zhoull@pku.edu.cn}
\author[pku,chep,chps]{Bo-Qiang Ma}\ead{mabq@pku.edu.cn}
\address[pku]{School of Physics and State Key Laboratory of Nuclear Physics and Technology,
\\Peking University, Beijing 100871, China}
\address[chep]{Center for High Energy Physics, Peking University, Beijing
100871, China} %
\address[chps]{Center for History and Philosophy of Science, Peking
University, Beijing 100871, China}%

\begin{abstract}
We make a reanalysis on the issue of neutrino speed anomaly by
taking into account the newly reported data from the ICARUS
experiment and other CNGS collaborations. We examine the consequence
of the Lorentz violation on the neutrino speed in a new framework of
standard model supplement (SMS), and find that the Lorentz violating
parameters are constrained at least one order stronger than that of
the earlier OPERA report. The combination with other
phenomenological considerations puts more stringent constraints on
the Lorentz violation of neutrinos.
\end{abstract}


\end{frontmatter}

\clearpage

\section{Introduction}
The neutrino speed anomaly reported by the OPERA
collaboration~\cite{opera11} attracted attentions on the
superluminality of neutrinos with various
speculations~\cite{review-on-opera}. It has been known that the
earlier OPERA release suffers from fatal hardware problem in time
measurement, and the measurement on neutrino velocity has been
updated~\cite{opera12}. Recently, the ICARUS
collaboration~\cite{ICARUS2012} released their new neutrino velocity
measurement with an earlier arrival time $\delta t = 0.10 \pm
0.67_{\mathrm{stat}} \pm 2.39_{\mathrm{sys}}$~ns with respect to the
expected traveling time of the light speed. The compatible time
deviations $\delta t$ are also reported by the Borexino and LVD
collaborations~\cite{AlvarezSanchez:2012wg,Agafonova:2012rh}. The
central value of the neutrino superluminality by the ICARUS
experiment is significantly reduced in magnitude, with a large error
compatible with either the superluminality or subluminality of
neutrinos. It is thus necessary to re-evaluate the issue of neutrino
speed anomaly from the new ICARUS result and the recent experiments
of the Borexino and LVD collaborations, combined with other
intuitive considerations in Refs.~\cite{Glashow11,Bi2011}.

Among many options for the realization of the theoretical Lorentz
violation (LV), we focus here on an
attempt 
to describe the LV effects based
on a basic principle that the equations describing the laws of
physics have the same form in all admissible mathematical
manifolds~\cite{mpla10,graal10,SMS3}. Such principle leads to the
following replacement of the ordinary partial $\partial_{\alpha}$
and the covariant derivative $D_{\alpha}$
\begin{equation}\label{eqn:substitution}
\partial^{\alpha} \rightarrow M^{\alpha\beta}\partial_{\beta},\quad
D^{\alpha}\rightarrow M^{\alpha\beta}D_{\beta},
\end{equation}
where $M^{\alpha\beta}$ is a local matrix by a splitting $M^{\alpha
\beta}=g^{\alpha \beta}+\Delta^{\alpha \beta}$. $g^{\alpha\beta}$ is
the metric of space-time and $\Delta^{\alpha \beta}$ is a new matrix
which brings new terms violating Lorentz invariance in the standard
model, therefore we denote the new framework as the Standard Model
Supplement (SMS)~\cite{mpla10,graal10,SMS3}. The magnitude of
Lorentz violation is determined by the matrix $\Delta^{\alpha
\beta}$, with the values of its elements to be measured or
constrained from experimental observations rather than from theories
at first.

We now extend the phenomenological applications of the SMS from the
cases of protons~\cite{mpla10} and photons~\cite{graal10,SMS3} to
the specific case of neutrinos. We examine the constraints on the LV
terms in the SMS by the available experiments concerning the
neutrino speed anomaly. We show the proportionality between the
neutrino superluminality and the neutrino LV parameters. Such
proportionality may serve to relate any possible neutrino speed
anomaly, actually unconfirmed yet, as possible hints for the Lorentz
violation of neutrinos.

\section{Lorentz violation of neutrinos}
For the sector of the electroweak interaction, the Lagrangian of
fermions in the SMS can be written as~\cite{mpla10}
\begin{eqnarray}\label{lagrangian_F}
\mathcal{L}_{\mathrm{F}} &=&
i\bar{\psi}_{A,\mathrm{L}}\gamma^{\alpha}\partial_{\alpha}\psi_{B,\mathrm{L}}\delta_{AB}+
i\Delta^{\alpha\beta}_{\mathrm{L},AB}\bar{\psi}_{A,\mathrm{L}}\gamma_{\alpha}\partial_{\beta}\psi_{B,\mathrm{L}}\nonumber\\
&&
+i\bar{\psi}_{A,\mathrm{R}}\gamma^{\alpha}\partial_{\alpha}\psi_{B,\mathrm{R}}\delta_{AB}+
i\Delta^{\alpha\beta}_{\mathrm{R},AB}\bar{\psi}_{A,\mathrm{R}}\gamma_{\alpha}\partial_{\beta}\psi_{B,\mathrm{R}},
~~~~
\end{eqnarray}
where $A,B$ are flavor indices. The LV terms are uniquely and
consistently determined from the standard model by the replacement
(\ref{eqn:substitution}). Generally, the LV matrix
$\Delta^{\alpha\beta}$ is particle-dependent~\cite{SMS3}, so it is
relevant to the flavors and has the flavor indices. For leptons,
$\psi_{A,\mathrm{L}}$ is a weak isodoublet, and
$\psi_{A,\mathrm{R}}$ is a weak isosinglet. After calculation of the
doublets and re-classification of the Lagrangian terms, the
Lagrangian can be written in a form like
Eq.~(\ref{lagrangian_F}) too. We assume that the LV matrix
$\Delta^{\alpha\beta}_{AB}$ is the same for fermions of
left-handedness and right-handedness, that is,
$\Delta^{\alpha\beta}_{\mathrm{L},AB}=\Delta^{\alpha\beta}_{\mathrm{R},AB}=\Delta^{\alpha\beta}_{AB}$.
When we do not consider mixing between the flavor $A$ and another
flavor $B$ for a given flavor $A$, we can rewrite
Eq.~(\ref{lagrangian_F}) as
\begin{equation}\label{lagrangian_F_total}
\mathcal{L}_{\mathrm{F}}=\bar{\psi}_A
(i\gamma^{\alpha}\partial_{\alpha}-m_A)\psi_A
+i\Delta^{\alpha\beta}_{AA}\bar{\psi}_A\gamma_{\alpha}\partial_{\beta}\psi_A,
\end{equation}
where $\psi_A=\psi_{A,\mathrm{L}}+\psi_{A,\mathrm{R}}$, i.e., the
field $\psi_A$ is the total contribution of the left-handed and
right-handed fermions of a given flavor $A$.  When there is only one
handedness for fermions, $\psi_A$ is just the contribution of this
handedness. We know that neutrinos are left-handed and antineutrinos
are right-handed in the standard model, therefore neutrino belongs
to the case of only one handedness. The Lagrangian $\mathcal{L}_{\mathrm{F}}$
contains also the mass terms. After calculations, we
can let $m_A\rightarrow 0$ for massless fermions. Then
$\partial{\mathcal{L}_{\mathrm{F}}}/\partial\bar{\psi}_A=0$ gives
the motion equation
\begin{equation}\label{motion_eqn}
(i\gamma^{\alpha}\partial_{\alpha}-m_A+i\Delta^{\alpha\beta}_{AA}\gamma_{\alpha}\partial_{\beta})\psi_A=0.
\end{equation}
This is also the modified Dirac equation, in which the LV term
$i\Delta^{\alpha\beta}_{AA}\gamma_{\alpha}\partial_{\beta}\psi_A$ is
determined by the LV matrix $\Delta^{\alpha\beta}_{AA}$ of fermions
with flavor $A$. Multiplying
$(i\gamma^{\alpha}\partial_{\alpha}+m_A+i\Delta^{\alpha\beta}_{AA}\gamma_{\alpha}\partial_{\beta})$
on both sides of Eq.~(\ref{motion_eqn}) and writing it in the
momentum space, we get the dispersion relation for fermions
\begin{equation}\label{disp_relation}
p^2+g_{\alpha\mu}\Delta^{\alpha\beta}_{AA}\Delta^{\mu\nu}_{AA}p_{\beta}p_{\nu}
+2\Delta^{\alpha\beta}_{AA}p_{\alpha}p_{\beta}-m_A^2=0.
\end{equation}
When we separate the spacial and temporal components of the 4-momentum
$p$, Eq.~(\ref{disp_relation}) reads
\begin{eqnarray}
&&(1+g_{\alpha\mu}\Delta^{\alpha 0}_{AA}\Delta^{\mu
0}_{AA}+2\Delta^{00}_{AA})E^2 \nonumber\\&+&
(2g_{\alpha\mu}\Delta^{\alpha 0}_{AA}\Delta^{\mu i}_{AA}+4\Delta^{(0i)}_{AA} )Ep_i\nonumber\\
&+&(g^{ij}+g_{\alpha\mu}\Delta^{\alpha i}_{AA}\Delta^{\mu
j}_{AA}+2\Delta^{ij}_{AA})p_ip_j-m_A^2=0,
\end{eqnarray}
which can be simplified as
\begin{equation}
\alpha E^2 +\alpha^i Ep_i +\alpha^{ij}p_ip_j-m_A^2=0,
\end{equation}
with the coefficients defined as
\begin{eqnarray}\label{lvm_combination}
\alpha &=& 1+g_{\alpha\mu}\Delta^{\alpha 0}_{AA}\Delta^{\mu 0}_{AA}+2\Delta^{00}_{AA}, \nonumber\\
\alpha^i &=& 2g_{\alpha\mu}\Delta^{\alpha 0}_{AA}\Delta^{\mu i}_{AA}+4\Delta^{(0i)}_{AA}, \nonumber\\
\alpha^{ij} &=& g^{ij}+g_{\alpha\mu}\Delta^{\alpha
i}_{AA}\Delta^{\mu j}_{AA}+2\Delta^{ij}_{AA},
\end{eqnarray}
in terms of the elements of the LV matrix. The velocity $v^i$ of the
fermion is the gradient of energy $E$ with respect to the momentum
$p_i$
\begin{equation}\label{velocity_component}
v^i\equiv {\partial E}/{\partial p_i}=-{(\alpha^i E
+2\alpha^{(ij)}p_j)}/{(2\alpha E + \alpha^i p_i)}.
\end{equation}
Then the magnitude of $v^i$ becomes
\begin{eqnarray}\label{velocity_magn}
&&v\equiv \sqrt{|v_iv^i|}=\sqrt{|g_{ij}v^iv^j|}\\
&=&\frac{1}{2\alpha E + \alpha^i p_i} \sqrt{|g_{ij}(\alpha^i E
+2\alpha^{(ih)}p_h)(\alpha^j E +2\alpha^{(jk)}p_k)|},\nonumber
\end{eqnarray}
in which the metric tensor
$g_{\alpha\beta}=\mathrm{diag}(1,-1,-1,-1)$ and the vacuum light
speed $c=1$.

All 16 degrees of freedom of the neutrino LV matrix are contained in
Eq.~(\ref{velocity_magn}). When parameterizing the 3-momentum $p_i$
with the spherical coordinate system, we shall see explicitly that
the velocity magnitude $v$ in Eq.~(\ref{velocity_magn}) is
direction-dependent, and this provides the possibility for an
anisotropy of the neutrino speed generally. We just focus on the
neutrino speed anomaly here, so we do not consider the
angle-dependence of $v$ and discuss only a specific form
$\Delta^{\alpha\beta}_{AA}=\mathrm{diag}(\eta,\xi,\xi,\xi)$ of the
SO(3) invariant LV matrix. Then Eq.~(\ref{velocity_magn}) becomes
\begin{equation}
v={(1-2\xi+\xi^2)}/{(1+2\eta+\eta^2)}\left|\vec{p}\right|/E,
\end{equation}
with the coefficients $\alpha=1+2\eta+\eta^2$, $\alpha^i=0$,
$\alpha^{ij}=(-1+2\xi-\xi^2)\delta^{ij}$ for
Eq.~(\ref{lvm_combination}), and the mass energy relation
\begin{displaymath}
E=\sqrt{((1-2\xi+\xi^2)\vec{p}^2+m_A^2)/(1+2\eta+\eta^2)}.
\end{displaymath}
So the deviation of the muon neutrino speed with respect to the
vacuum light speed is
\begin{eqnarray}
\delta v&\equiv&{(v-c)}/{c}=-(\eta+\xi)-({1}/{2})\left({m_A}/{E}\right)^2\nonumber\\
&=&-(\eta+\xi), \quad m_A\ll E,\label{neu_speed_anomaly}
\end{eqnarray}
where $c=1$. Eq.~(\ref{neu_speed_anomaly}) shows clearly that the
neutrino superluminality is related to the LV parameters directly.
Neutrinos could be either superluminal or subluminal according to
the signs and magnitudes of their LV parameters. The early work by
Coleman and Glashow in Ref.~\cite{Coleman99} proposed the original
thought on the particle velocity anomaly due to LV. The mass energy
relation Eq.~(17) in Ref.~\cite{mpla10} for the proton means that
the fermion velocity $v$ can be larger or less than the vacuum light
speed $c$, and that the difference ${(v-c)}/{c}$ is proportional to
the parameter $\xi$ in the corresponding LV matrix for protons. As
for the massless gauge bosons, the difference between the photon
propagating velocity $c_{\gamma}$ and the Lorentz invariant light
speed $c$ is proportional to the elements of the LV matrix of photons
too~\cite{graal10}, i.e., $\delta c_{\gamma}\equiv
{(c_{\gamma}-c)}/{c}\propto \xi$, where $\xi$ corresponds now to the
LV parameter for photons.

\section{Comparison with the Coleman-Glashow model}
We now compare our results of the LV effects in the neutrino sector
from the SMS with that of the Coleman-Glashow model in
Ref.~\cite{Coleman99}, which is a simple and intuitive model to
include LV terms for high energy particles. When the LV matrix
$\Delta_{AA}^{\alpha\beta}$ is diagonal, e.g.,
$\Delta_{AA}^{\alpha\beta}=\textrm{diag}(\eta,\xi,\xi,\xi)$,
Eq.~(\ref{disp_relation}) reads
\begin{eqnarray}
E^2&=&(|\vec{p}|^2(1-2\xi+\xi^2)+m_A^2)/(1+2\eta+\eta^2)\nonumber\\
&=&|\vec{p}|^2 c_A^2+(m_A^{'})^2c^4_A.\label{Coleman_G_disp}
\end{eqnarray}
With a reformulation, the dispersion relation of
Eq.~(\ref{Coleman_G_disp}) can be rewritten as
\begin{equation}\label{Coleman_G_disp2}
p^2-(m_A^{'}c^2_A)^2=\epsilon \vec{p}^2,
\end{equation}
where $c_A^2=1+\epsilon$ with $\epsilon=-2(\eta+\xi)$ (cf. Eq.~(2.18) in Ref.~\cite{Coleman99}), 
and $m_A^{'}$ is a redefinition of $m_A$. $c_A$ is called the
maximal attainable velocity of type $A$ particles in
Refs.~\cite{Coleman99,Glashow11}. Hence, the maximal attainable
velocity of particles in the Coleman-Glashow model corresponds to
a specific case of a diagonal LV matrix
$\Delta^{\alpha\beta}$ in the SMS framework.

The LV term $\epsilon \vec{p}^2$ in Eq.~(\ref{Coleman_G_disp2})
corresponds to the contribution from a LV term $\partial_i\Psi
\epsilon \partial^i\Psi$ added to the standard Lagrangian for a
field $\Psi$~\cite{Coleman99}. The term like $\partial_i\Psi
\epsilon\partial^i\Psi$ can incorporate the essence of LV
effects intuitively, that is, the Lorentz violations reflect
influences of some unknown background fields (i.e., $\epsilon$ here)
to the standard matter fields under our attentions. So the formula
Eq.~(\ref{Coleman_G_disp2}) can grasp the essential effects of
the Lorentz violation, though $\epsilon$ there does not have four
spacetime indices and is defined in some particular observer frame
therefore. Some more detailed discussions about this model and the
corresponding three scenarios of Lorentz violation are provided in
Ref.~\cite{review-on-opera}. Since the effect of $\epsilon$ can be
equivalently represented by $\Delta^{\alpha\beta}$s, we adopt the
latter formalism in our phenomenological analysis for generality.
Sometimes, the relations of $\epsilon$ in the Coleman-Glashow model
and $\Delta^{\alpha\beta}$s here can be used to check the
consistency of the calculated results.

\section{The neutrino speed anomaly in data}

We now re-evaluate the constraints on LV parameters according to the
ICARUS result~\cite{ICARUS2012} and the data of other CNGS collaborations~\cite{AlvarezSanchez:2012wg,Agafonova:2012rh},
combined with previous data~\cite{MINOS07,Longo87} relevant to possible neutrino speed
anomaly.

{\bf ICARUS}: The ICARUS experiment reported the new
neutrino velocity of the CNGS neutrino beam~\cite{ICARUS2012}. The
flying distance of muon neutrinos from CERN to ICURAS neutrino
detector is around $730$~km. The arrival time of the flying
neutrinos is earlier than that expected with the light speed with a
small measured value $\delta
t=0.10\pm0.67_{stat}\pm2.39_{sys}$~ns.
The corresponding superluminality $\delta v$ is calculated
\begin{displaymath}
\delta v=\left(0.4\pm2.8_{stat}\pm9.8_{sys}\right)\times 10^{-7},
\end{displaymath}
where the central value of $\delta v$ is smaller than that of the
OPERA report by two orders in magnitude, with a large error bar that
is compatible with either superluminality or subluminality of
neutrinos. The corresponding time deviation $\delta t$ of the CNGS
beam with respect to the expected light speed  has been reported by
the Borexino and LVD collaborations respectively in
Refs.~\cite{AlvarezSanchez:2012wg,Agafonova:2012rh}. $\delta
t=-0.8\pm0.7_{stat}\pm2.9_{sys}$ ns (Borexino). $\delta
t=0.3\pm0.6_{stat}\pm3.2_{sys}$ ns (LVD). The relevant neutrino
speed of the recent CNGS experiments is listed in
Tab.~\ref{tab_neutrino_speed}, where it is shown that all the
superluminallity of neutrinos of the CNGS beam is consistent with
zero.
\begin{table}
  \centering
  \caption{Neutrino velocity of CNGS beam. Total standard deviation is used for weighted average:
  $\sigma=\sqrt{\sigma_{stat}^2+\sigma_{sys}^2}$. }\label{tab_neutrino_speed}
\begin{tabular}{cc}
  \hline
  Collaboration & Superluminality $\delta v$\\
  \hline
  OPERA & $\left(2.7\pm 3.1_{stat} {}^{+3.4}_{-3.3}{}_{sys}\right)\times 10^{-6}$ \\
  ICARUS & $\left(0.4\pm 2.8_{stat}\pm 9.8_{sys}\right)\times 10^{-7}$ \\
  Borexino & $\left(-3.3\pm 2.9_{stat}\pm 11.9_{sys}\right)\times 10^{-7}$ \\
  LVD & $\left(1.2\pm 2.5_{stat} \pm 13.2_{sys}\right)\times 10^{-7}$ \\
  \hline
  Weighted average & $\left(0.06\pm6.7\right)\times 10^{-7}$ \\
  \hline
\end{tabular}
\end{table}

{\bf MINOS and SN1987a}: The velocity of $\sim$3~GeV muon
neutrinos of the MINOS detectors was reported in
Ref.~\cite{MINOS07}. The value of neutrino superluminality is
$\delta v=(5.1\pm 2.9)\times 10^{-5}$. The electron neutrinos of the
observation SN1987a also put a bound on the deviation of the
velocity $v_{\nu_e}$ of neutrinos (actually anti-neutrinos
${\bar{\nu}_e}$) with respect to the light speed $c$. The constraint
on the neutrino superluminality is $|\delta v|\leq 2\times 10^{-9}$
from Ref.~\cite{Longo87}.

The energy dependent term in Eq.~(\ref{neu_speed_anomaly}) is not
able to provide a consistent explanation of different magnitudes of
neutrino speed anomaly in different data. The energy of the
CNGS muon neutrino is $\sim17$~GeV and the neutrino mass is
estimated to be less than 1~eV up to now. Around the energy range of
10~MeV for SN1987a~\cite{SN1987a}, even if we take the mass of
electron neutrinos to be of order 1~eV~\cite{Weinheimer99}, the mass
term $m_{\nu_e}^2/(2E^2)$ in Eq.~(\ref{neu_speed_anomaly}) is $\sim
10^{-14}$, which can not produce the speed anomaly of order
$10^{-9}$ if there exists speed anomaly for SN1987a. So, if there is
indeed Lorentz violation as the central values of the data suggest,
there are two possibilities: (i) The Lorentz violation for neutrinos
of three generations are the same, but the Lorentz violation should
be constrained to be very small to be consistent with the nearly zero
speed anomaly in the data; (ii) The Lorentz violation of
neutrinos is generation-dependent. Possibility (ii) is not discussed
here. We just consider the former one in the analysis.

The CNGS experiments give a averaged constraint $\sim 10^{-7}$ (1$\sigma$) on
the superluminality. Given the more stringent result of the SN1987a,
the experiments put a constraint on the superluminality
$\delta v$ of neutrinos, that is,
\begin{equation}\label{speed_contraint}
\delta v_{\nu} <1\times 10^{-9}.
\end{equation}
With Eq.~(\ref{neu_speed_anomaly}), the constraint on the speed
anomaly of neutrinos means that the LV parameters $\eta_{\nu}$ and
$\xi_{\nu}$ satisfy
\begin{equation}\label{LV_constraint}
|\eta_{\nu}+\xi_{\nu}| < 1\times 10^{-9}.
\end{equation}
More experiments will provide more details on Lorentz violation of
neutrinos. In previous analyses, we get the LV parameter
$|\xi_p|\leq 10^{-23}$ for high energy protons~\cite{mpla10} and
$|\xi_{\gamma}|\leq 10^{-14}$ for photons~\cite{graal10} by
confronting with relevant experimental observations. The constraint
on the neutrino LV parameters is weaker than those of protons and
photons.

Other neutrino experiments or observations may constrain the
superluminality and Lorentz violation of neutrinos stronger or
weaker than Eqs.~(\ref{speed_contraint}) and~(\ref{LV_constraint}).
As Cohen and Glashow revealed~\cite{Glashow11}, getting rid of the
possible Cherenkov analogous radiations of neutrinos for reaching
the OPERA detector, there should be a constraint $\delta v_{\nu}
<(2m_e/E_{\nu_{\mu}})^2/2=1.8\times 10^{-9}$ for OPERA muon
neutrinos of $\sim 17$~GeV. A stronger constraint is also provided
in Ref.~\cite{Glashow11}, that is, $\delta v_{\nu}<0.85\times
10^{-11}$ (corresponding to $\delta < 1.7\times 10^{-11}$ there)
from observations of very high energy neutrino events. Based on the
production process (kaon decay) of muon neutrinos, the constraint
$\delta v_{\nu}< 3\times 10^{-7}$ is required such that muon
neutrinos can be produced~\cite{Bi2011}. It is necessary to notice
that although the non-velocity experiments or observations can
constrain also the superluminality or Lorentz violation weakly or
strongly, the interpretations for experimental mechanisms are still
model sensitive. Then the derived constraints from these experiments
should be treated more prudently than the results of neutrino
velocity experiments by measuring directly flying length and time.
So the constraints of Eqs.~(\ref{speed_contraint})
and~(\ref{LV_constraint}) are still kept in the article.

The constraints of Eqs.~(\ref{speed_contraint})
and~(\ref{LV_constraint}) indicate also the necessary precision for future
experimental facilities to measure the neutrino superluminality on Earth. For
the flying distance $L=730$~km of the neutrino of the CNGS beam, the precision
of the corresponding GPS of time measurement should be smaller than
$2.43\times10^{-3}$~ns, which is beyond the precision of
current GPS technology. Even if we consider the maximally
accumulated consequences of the Lorentz violation with neutrinos
flying right through Earth, that is, with $L$ being the diameter of
Earth, the time difference $\delta t$ of neutrinos with respect to
the light speed is $\delta t < 4.25\times 10^{-2}$~ns. So, in order
to test the neutrino Lorentz violation of an order smaller than
$10^{-9}$ or constrain it more stringently, it is a big challenge to the experimental technology for
neutrino velocity measurements on Earth, combined with also the
uncertainties in length measurement.

The following are some relevant remarks.
We can compare the constraints on the LV parameters from the
neutrino velocity experiments with those of attributing neutrino
oscillations to resulting purely from the Lorentz violation in previous
analyses~\cite{Xiao08,Yang09}. In the model of Lorentz violation for
neutrino oscillations, neutrinos can be taken as massless and
neutrino flavor states are mixing states of energy eigenstates. As a
consequence of Lorentz violation, different flavor states of
neutrinos mix with each other when neutrinos of different
eigenenergies propagate in space. From Ref.~\cite{mpla10}, the
complete Lagrangian $\mathcal{L}_{\mathrm{F}}$ of fermions (cf.
Eq.~(\ref{lagrangian_F_total})) is
\begin{eqnarray}\label{larangian_F_oscillation}
&&\mathcal{L}_{\mathrm{F}}=\bar{\psi}_A (i\gamma^{\alpha}\partial_{\alpha}-m_A)\psi_B\delta_{AB}\nonumber \\
&&+i\Delta^{\alpha\beta}_{AB}\bar{\psi}_A\gamma_{\alpha}\partial_{\beta}\psi_B
-g\Delta^{\alpha\beta}_{AB}\bar{\psi}_A
\gamma_{\alpha}A_{\beta}\psi_B,
\end{eqnarray}
where $A,B,\cdots$ are the flavor indices, $A_{\beta}$ is the gauge
bosons, and $g$ is the coupling constant. In the case of massless
neutrinos, we can let $m_A\rightarrow 0$ for the derived results.
From the viewpoint of the effective field theory, the vacuum
expectation value $\langle
g\Delta^{\alpha\beta}_{AB}A_{\beta}\rangle$ can be treated
equivalently as another LV parameter $a^{\alpha}_{AB}$, so
Eq.~(\ref{larangian_F_oscillation}) reads also
\begin{eqnarray}
&& \mathcal{L}_{\mathrm{F}}=\bar{\psi}_A
(i\gamma^{\alpha}\partial_{\alpha}-m_A)\psi_B\delta_{AB}
\nonumber\\&&
+i\Delta^{\alpha\beta}_{AB}\bar{\psi}_A\gamma_{\alpha}\partial_{\beta}\psi_B
-a^{\alpha}_{AB}\bar{\psi}_A \gamma_{\alpha}\psi_B.\nonumber
\end{eqnarray}
The derivative
$\partial\mathcal{L}_{\mathrm{F}}/\partial\bar{\psi}_A$ leads to the
motion equation
\begin{displaymath}
(i\gamma^{\alpha}\partial_{\alpha}-m_A)\psi_B\delta_{AB}
+i\Delta^{\alpha\beta}_{AB}\gamma_{\alpha}\partial_{\beta}\psi_B
-a^{\alpha}_{AB}\gamma_{\alpha}\psi_B=0.
\end{displaymath}
By multiplying $\gamma^0$ on both sides, the terms corresponding to
the operator $i\partial_t$ are the Hamiltonian. So the
Hamiltonian $\mathcal{H}_{AB}$ is
\begin{displaymath}
\mathcal{H}_{AB}= -\gamma^0(i\gamma^{k}\partial_{k}-m_A)\delta_{AB}
-i\Delta^{\alpha\beta}_{AB}\gamma^0\gamma_{\alpha}\partial_{\beta}
+a^{\alpha}_{AB}\gamma^0\gamma_{\alpha}.
\end{displaymath}
All following derivations have been done in Ref.~\cite{Yang09},
including the reductions of the Hamiltonian to the level of quantum mechanics
and the solutions of the energy eigenstates. It is found that
even with very small LV parameters, $\Delta^{00}_{\nu_\mu
\nu_\mu}\simeq 10^{-20}$, one can still explain the neutrino
oscillation in experiments by pure LV effects. The parameter
$\Delta^{00}_{\nu_\mu \nu_\mu}$ is denoted by $c^{00}_{\mu \mu}$
and $c^{00}_{\mu \mu}\simeq 10^{-20}$ in Ref.~\cite{Yang09}. The
parameter $\Delta^{00}_{\nu_\mu \nu_\mu}$ is of order
$10^{-20}$, when the neutrino oscillations are fitted using the LV
parameters. In this paper, $\Delta^{00}_{AA}$ is just written as
$\eta$, so we find that $\eta_{\nu_\mu}\simeq 10^{-20}$. Based on
Eq.~(\ref{LV_constraint}), $|\xi_{\nu_\mu}|< 10^{-9}$. So, if there
is Lorentz violation (even tiny) of neutrinos, it can be possible
that $|\xi_{\nu_\mu}|\gg|\eta_{\nu_\mu}|$, and that the spacial Lorentz
violation is much larger than the temporal one of the muon neutrino,
since the parameter $\xi_{\nu_\mu}$
provides contributions to the neutrino speed anomaly from Lorentz
violation and it belongs to the spacial part of the LV matrix of
muon neutrinos. At this stage, it is not known whether the spacial parts of
LV are larger than the temporal ones indeed in Nature if the LV exists.

Nevertheless, it should be noted also that Ref.~\cite{Yang09}
included 3 specific forms of the tensors  $a^{\mu}_{AB}$ and
$\Delta^{\alpha\beta}_{AB}$, whereas it is assumed here that
$\Delta^{\alpha\beta}_{AB}$ is diagonal for confronting with the
formalism in the Coleman-Glashow model. The consistency between the
results of neutrino speed anomaly and the neutrino oscillation
experiments can be further studied with a more general form of the
LV matrix. There are large degrees of freedom in
$\Delta^{\alpha\beta}_{AB}$, and it is still possible both the
Lorentz violation and the conventional oscillation mechanism with
neutrino masses contribute to neutrino oscillations. In that case,
there are totally $16\times 6+3=99$ degrees of freedom, i.e., 6
matrices $\Delta^{\alpha\beta}_{AB}$ for 3 kinds of neutrinos (when
$\Delta^{\alpha\beta}_{AB}$ has the symmetric indices $A$ and $B$)
together with 3 neutrino masses within the SMS framework, to adjust
parameters for confronting with relevant experiments. The option of
particle and antiparticle difference could also double the allowed
degrees of freedom. This provides a more compelling expectation to
confront with experiments.

\section{Conclusion}

The speed anomaly of the neutrino experiments, such as
the CNGS experiments and etc., is compatible with zero,
and serves on the other hand as stringent constraints
on the Lorentz violation of neutrinos, although the possible
superluminality of the available data can be explained
by the corresponding Lorentz violation of the neutrino sector if Lorentz violation
exists indeed.
At the moment, both the superluminality and
the subluminality of neutrinos are permitted by data and by theory.
Therefore the superluminality and Lorentz violation of neutrinos still needs to be
examined or ruled out by further experiments. The experiments on
neutrino speed might provide the chance to reveal novel neutrino
properties beyond conventional understandings.

\section*{Acknowledgments}
The work is supported by National Natural Science Foundation of
China (Nos. 10975003, 11021092, 11035003, and 11120101004), by
the Key Grant Project of Chinese Ministry of Education (No. 305001),
and by the Research Fund for the Doctoral Program of Higher
Education, China.

\end{document}